# Building up user confidence for the spaceborne derived global and continental land cover products for the Mediterranean region: the case of Thessaly

**Ioannis Manakos[1*], Christina Karakizi[2], Giannis Gkinis[2] and Konstantinos Karantzalos[2]**

[1] Information Technologies Institute, Centre for Research and Technology Hellas, Greece
[2] Remote Sensing Laboratory, National Technical University of Athens, Greece; chrkarakizi@central.ntua.gr (C.K.); giannisginis53@gmail.com (G.G.); karank@central.ntua.gr (K.K.)
**\*** Correspondence: imanakos@iti.gr (I.M.); Tel.: +30-2311-257760 (I.M.)

**Abstract:** Across globe and space agencies nations recognize the importance of homogenized land cover information, prone to regular updates, both in the context of thematic and spatial resolutions. Recent sensor advances and the free distribution policy promote the utilization of spaceborne products in an unprecedented pace into an increasingly wider range of applications. Ensuring credibility to the users is a major enabler in this process. To this end this study contributes with a systematic accuracy performance measurement and continental/global land cover layers' inter-comparison moving towards confidence built up. Confidence levels during validation and a weighted overall accuracy assessment were applied. Google Earth imagery was employed to assess the accuracy of three land cover products, i.e., Globeland30, HRLs and CLC 2012, for the years 2010 and 2012. Reported rates indicate a minimum weighted overall accuracy of 84%. Specific classes' performance deviations from the general trend were noted and discussed on the basis of an unbiased sampling approach. By integrating confidence levels during the ground truth annotation, stratified sampling on the several Corine Level 3 subclasses and the weighted overall accuracy assessment, the different aspects of the considered land cover products can be highlighted more objectively.

## 1. Introduction

Natural and ecological processes reach out of the human-induced limited space, as delineated by the administrative boundaries, demanding for standardized products, beyond locally generated land cover products, to feed in models and scenarios. For example, tele-couplings (e.g. in the form of large area acquisitions or climatic changes) are discussed and measured across the globe for their consequences in the land use and the local societies along with its value return to the global market [1]. Activities on land and sea are increasingly depended by frequently updated qualitative land cover products.

Recent advances in data provision frequency and accessibility by the global scientific community, the progress in Earth Observation techniques and big data handling, enabled the generation of numerous continental and global land cover products (C/GLC) with increasing spatial resolution and frequency. Issues and challenges accompany these developments, mainly in matters of data interpretation and categorization, as well as surface objects' delineation and exact location.

Compatibility and interpretation issues are already being treated by working groups, such as the European Environment Information and Observation Network (EIONET) Action Group on Land monitoring in Europe (EAGLE). At the same time globally performed exercises and fora are initiated from the C/GLC producers themselves and in coordination with international remote sensing Associations in order to locally and regionally validate the land cover products. The latter occurs as a necessity to account for the products' dependence on a huge variety of geographical and climatic conditions and enhance credibility towards policy makers, stakeholders and entrepreneurs. Confidence needs to be built up.

C/GLC maps represent the most important sources of accumulative and homogenized information about the surface of the earth and are used for several policies and scientific applications such as environmental monitoring, water monitoring, biodiversity, urban planning and change detection of global land cover [2-4]. There are several C/GLC maps, such as IGBP-DISCover, GlobCover maps, MODIS GLC, LC-CCI maps and FROM-GLC maps [3]. Currently, many organizations produce C/GLC maps with higher resolution, namely the Land Cover-CCI (LC-CCI) maps at 300m, GIO High Resolution Layers at 20m and Globeland at 30m [3, 5, 6]. They all have been produced by remote sensing analysis using various optical data and methods. However, they are produced as independent datasets with different class hierarchies, semantic class similarities and as a consequence considerable disagreement among them have been reported [2, 3, 7].

Comparative accuracy assessment of C/GLC maps, either against to one another or juxtaposed against very high resolution ground data, is crucial but challenging, because of the lack of reference data. Several studies have assessed C/GLC products to analyze their weaknesses and strengths [2, 8-10]. A few studies compared the accuracy estimates by harmonizing confusion matrices, but it remains unclear how they compared the C/GLC maps with the same reference dataset [2, 3, 7]. Reference datasets that are suitable for multiple maps were developed and used for validation of C/GLC maps [11]. However, these studies provide spatial agreement between C/GLC maps, but they neither compare the very recent high resolution products nor they estimate confidence levels during validation.

Triggered by the aforementioned challenges and recent developments this study presents an approach to assess the confidence a user shall have to recently produced C/GLCs. In particular, a comparative accuracy assessment was performed between different C/GLCs. The confidence levels of the expert were incorporated during the validation through a weighted overall accuracy assessment during the evaluation and inter-comparison. The focus was on a representative landscape of the Northern Mediterranean basin, the area of Thessaly in Greece. Existing Google Earth images were employed to assess the accuracy of C/GLCs, namely CORINE Land Cover 2012, GIO High Resolution Layers and Globeland30, for the years 2010 and 2012. In addition, the type of reported errors among the semantic classes is discussed revealing further qualitative aspects of the considered C/GLCs.

**2. Materials and Methods**

*2.1. Study Area*

The study area is centered at 39°24'8.06"N latitude and 21°59'1.10"E longitude (Figure 1). It is located in central Greece near the regions of Macedonia, Epirus, Central Greece and borders with the Aegean Sea on the east. It covers an area of 14,036 square kilometers (Fig. 1), and includes (since *Kallikratis* reform of 2010) five Local Administrative Units (LAU) of 2nd level (former NUTS 3 - Nomenclature of Territorial Units for Statistics) and 25 municipalities. It also includes the Sporades Islands. Thessaly can be considered an optimal region for validating land cover mapping results, since it presents high landscape and land cover diversity. The overall topography of the area consists of mountainous and semi-mountainous areas on the perimeter and lowlands in the center. The whole area includes five mountains, containing the spur of Pindus that peaks up to Mount Olympus, with an altitude of 2917 m above sea level. The area comprises additionally of rural and forested areas, with the Thessaly plain to be one of the most important agricultural areas of the country. The rest of the land cover is divided into urban areas and ranching land [12]. Moreover, Thessaly contains certain protected Natura2000 areas (e.g. Lake Karla), several statutory and non-protected areas, as well as one UNESCO World Heritage Site, namely Meteora [12].

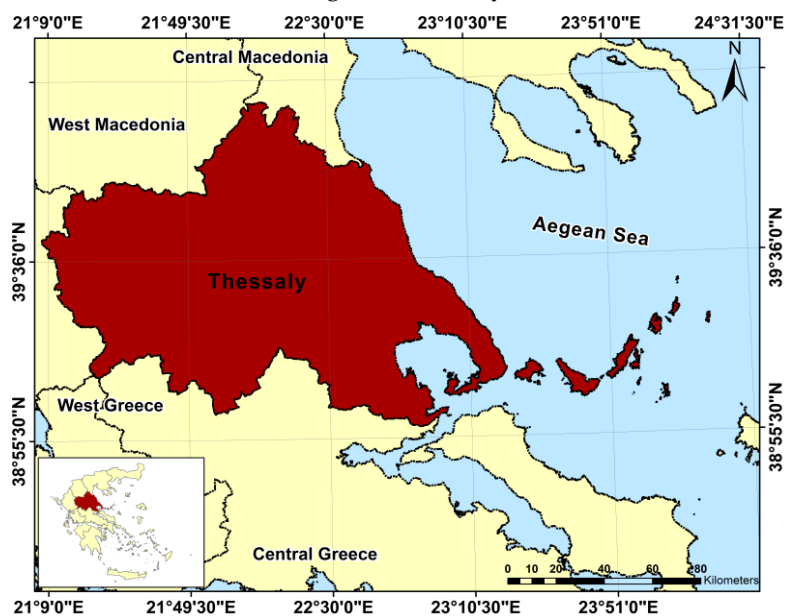

**Figure 1.** The study area: Thessaly is located in central Greece and presents high landscape and land cover diversity.

*2.2. Data Sets*

In this study, three openly available LC maps are employed, namely the Copernicus CORINE Land Cover 2012 (CLC2012) map [13], the Pan-European Copernicus GIO High Resolution Layers (HRLs) of 2012 [14] and the GlobeLand30 [5]. In particular, CLC2012 represents the 4th CORINE Land Cover inventory. For its production a dual coverage of satellite images (IRS P6 LISS III and RapidEye) was used. It consists of an inventory of land cover in 44 classes. Corine Land Cover 2012 products are available in both raster (100m and 250m resolution), and vector (ESRI and SQLite geodatabase) formats in the European projection system (EPSG: 3035). Pan-European High Resolution Layers (HRLs) are designed as complementary to land cover/land use mapping, such as in the CORINE land cover (CLC) datasets [15]. The HRLs are produced from 20 m resolution satellite imagery through a combination of automatic processing and interactive rule based classification. The data is available as georeferenced raster data in the European projection (EPSG: 3035) with 20m and 100m spatial resolution. HRLs consist of five thematic classes corresponding to main thematic classes of the CLC. GlobeLand30 (GLC30) dataset production involved multispectral images with 30 meters spatial resolution, including the TM5 and ETM+ of America Land Resources Satellite (Landsat) and the multispectral images of China Environmental Disaster Alleviation Satellite (HJ-1). The classification system includes 10 land cover types and uses as a baseline the year 2010 [5]. The data is available as georeferenced raster data in WGS 84 coordinate system, UTM projection, 6-degree zoning and reference ellipsoid WGS 84 ellipsoid, with 30m resolution [16].

During the validation process, the main objective was the inter-comparison of four different LC categories, namely the *(i) Artificial Surfaces, (ii) Forest, (iii) Water,* and *(iv) Agriculture*. These four LC classes are related with the *Artificial Surfaces, Forest and Semi-Natural Areas* (including only the Level-2 class *Forests*), *Water Bodies* (including only the Level-2 subclass of *Inland Waters*) and *Agricultural Areas* from the CLC2012 dataset. Regarding the HRLs dataset, the 20m layers of *Imperviousness, Forest Type,* and *Permanent Water Bodies* were only employed, since HRLs do not offer a specific layer for agriculture. Regarding GlobeLand30, the classes *Artificial Surfaces, Forest, Water Bodies,* and *Cultivated Land* were studied. To this end, Table 1 indicates the class/name correspondences among the different selected LC classes, while Table 2 demonstrates in detail the CLC 2012 nomenclature from Level 1 to Level 3 for the classes and sub-classes studied.

Table 1. The corresponding LC classes between the different employed LC maps.

| General Categories | GLC30 2010 - 30m | HRLs 2012 - 20m | CLC2012 Level 1 - 100m |
|---|---|---|---|
| Artificial Surfaces | Artificial Surfaces (code 80) | Imperviousness | 1. Artificial Surfaces |
| Forest | Forest (code 20) | Forest Type | 3. Forest and Semi-Natural Areas |
| Water | Water Bodies (code 60) | Permanent Water Bodies | 5. Water Bodies |
| Agriculture | Cultivated Land (code 10) | - | 2. Agricultural Areas |

Table 2. CLC 2012 class nomenclature from Level 1 to Level 3 for the four studied land cover categories.

| Corine Land Cover 2012 Levels | | |
|---|---|---|
| **Level 1** | **Level 2** | **Level 3** |
| 1. Artificial Surfaces | 1.1. Urban fabric<br>1.2. Industrial, commercial and transport units<br>1.3. Mine, dump and construction sites<br>1.4. Artificial, non-agricultural vegetated areas | 1.1.1. Continuous urban fabric<br>1.1.2. Discontinuous urban fabric<br>1.2.1. Industrial or commercial units<br>1.2.2. Road and rail networks and associated land<br>1.2.3. Port areas<br>1.2.4. Airports<br>1.3.1. Mineral extraction sites<br>1.3.2. Dump sites<br>1.3.3. Construction sites<br>1.4.1. Green urban areas<br>1.4.2. Sport and leisure facilities |
| 3. Forest and Semi-Natural Areas | 3.1. Forests | 3.1.1. Broad-leaved forest<br>3.1.2. Coniferous forest<br>3.1.3. Mixed forest |
| 5. Water Bodies | 5.1. Inland waters | 5.1.1. Water courses<br>5.1.2. Water bodies |
| 2. Agricultural Areas | 2.1. Arable land<br>2.2. Permanent crops<br>2.3. Pastures<br>2.4. Heterogeneous agricultural areas | 2.1.1. Non-irrigated arable land<br>2.1.2. Permanently irrigated land<br>2.1.3. Rice fields<br>2.2.1. Vineyards<br>2.2.2. Fruit trees and berry plantations<br>2.2.3. Olive groves<br>2.3.1. Pastures<br>2.4.1. Annual crops associated with permanent crops<br>2.4.2. Complex cultivation patterns<br>2.4.3. Land principally occupied by agriculture, with significant areas of natural vegetation<br>2.4.4. Agro-forestry areas |

*2.3. Inter-Comparison and Validation Methodology*

2.3.1. Sampling design based on L3 CLC2012

The accuracy assessment of the three different C/GLC maps was initially performed based on the inter-comparison with samples, which were derived after a comprehensive sampling procedure based on the CLC2012. In particular, random sampling points were generated automatically for each CLC2012 Level 3 subclass on the vector files under a stratified sampling design. These sampling points (X,Y) were then assigned to the nearest pixel (i.e. nearest pixel's center) for HRLs and GlobeLand30 land cover product (see Section 3.3.3).

The sampling unit was chosen to be one pixel for each land cover product since samples larger than the minimum mapping unit are frequently correlated with complicated cases of mixed samples, including more than one land-cover type [17]. Regarding the decision over the size of the sample that forms the testing set we adopted the basic sampling theory to derive an estimate of the required sample size. More analytically we apply the approach proposed for defining the testing set in remote sensing studies [18], using the equation:

$$n = \frac{z_{a/2}^2 \, P(1-P)}{h^2} \quad (1)$$

where $n$ is the sample size, $z_{a/2}$ the critical value of the normal distribution for the two-tailed significance level $\alpha$, $P$ is a planning value for the correctly allocated cases population proportion and $h$ the half width of the desired confidence interval. Here we consider a typically adopted 0.05 significance level giving a $z_{a/2}$ equal to 1.96. We also use a large conservative value for $P$ of 0.5 and we aim at having a confidence interval between ±4% up to ±5%. For a $h$ of 0.04 applying the eq. 1 we get an estimation of sample size of 601 samples and for the value 0.05 we get 385 samples.

A stratified sampling scheme [3, 18-21] was employed based on the area proportion of each L3 subclass to the total cover area, while applying constrains over maximum and minimum sizes for the subclasses. In order to remain into the required $n$ range of 385-601 samples we set a maximum of 120 and a minimum of 5 per L3 class. More analytically, CLC2012 class covering the highest relative extent for CLC2012 in the study area i.e., 2.1.2 *Permanently irrigated land* covering a 25.03%, is given the maximum of 120 samples. All other Level 3 categories are attributed a sample number proportionally to the above relationship and their relative extent. However, the minimum of 5 samples is given to each CLC 2012 L3 category, that was attributed with less than five samples, resulting to an overall sample size of 539 samples (inside the required range). Table 3 presents the percentage of the relative extent of each subclass and the derived number of samples, based on the procedure described above.

2.3.2. HRLs and GlobeLand30 Data Pre-Processing

Standard data preprocessing techniques were required for a more efficient data/map management and direct comparison. Both maps were re-projected into the same reference system (CRS), which was the CRS of the two Copernicus datasets; i.e. ETRS89 / ETRS-LAEA. Basic image algebra techniques were applied for multiplying the given HRLs layers with different constant numbers so as, after the merging of all the layers, each LC class to have a unique value. The main goal here was to create a single raster file/map for each one of the HRLs and GLC30 datasets. Regarding the GLC30 dataset, the three raster files (i.e. n34_35_2010lc030.tif, n34_40_2010lc030.tif, n35_35_2010lc030.tif.) that contained the area of interest with all the classes and subclasses were mosaicked, and then the output file was re-projected from WGS 84 /UTM zone 34-35N to ETRS89/ ETRS-LAEA. It should be noted that a geolocation (mostly translation) disagreement of about 0.5-1 pixel was observed between the three different GLC30 raster files. At the same time, by optically comparing the given products at the coastline areas, file n34_40_2010lc030.tif was in better geolocation agreement with the Copernicus datasets (CLC2012, HRLs) than the other two raster files; thus, the GLC30 TIF files disagreement was handled during the merging procedure by shifting the two remaining raster files based on the n34_40_2010lc030.tif. All in all, the final merged products for the three C/GLC datasets presented a mis-registration disagreement of about 0-3 HRLs 20m pixels in comparison with the coastline delineated with the images present in Google Earth.

**Table 3.** The number of samples that were randomly collected per L3 CLC2012 subclass based on a stratified sampling methodology. Overall, more than 500 samples were collected, *i.e.*, 62 for Artificial Surfaces, 338 for Agriculture, 129 for Forest and 10 for the Water class.

| CLC L1 Classes | L3 Sub-Classes | Area Coverage | # of samples per coverage | Selected # of samples |
|---|---|---|---|---|
| | **CLC2012-based Stratified Sampling** | | | |
| Artificial Surfaces | 1.1.1. Continuous urban fabric | 0.02% | 0.07 | 5 |
| | 1.1.2. Discontinuous urban fabric | 2.53% | 12.12 | 12 |
| | 1.2.1. Industrial or commercial units | 0.50% | 2.41 | 5 |
| | 1.2.2. Road and rail networks and associated land | 0.19% | 0.89 | 5 |
| | 1.2.3. Port areas | 0.00% | 0.02 | 5 |
| | 1.2.4. Airports | 0.26% | 1.25 | 5 |
| | 1.3.1. Mineral extraction sites | 0.12% | 0.55 | 5 |
| | 1.3.2. Dump sites | 0.00% | 0.01 | 5 |
| | 1.3.3. Construction sites | 0.03% | 0.15 | 5 |
| | 1.4.1. Green urban areas | 0.00% | 0.02 | 5 |
| | 1.4.2. Sport and leisure facilities | 0.03% | 0.16 | 5 |
| | | | *Sum Artificial Surfaces:* | **62** |
| Agriculture | 2.1.1. Non-irrigated arable land | 24.00% | 115.02 | 115 |
| | 2.1.2. Permanently irrigated land | 25.03% | 120.00 | 120 |
| | 2.1.3. Rice fields | 0.01% | 0.07 | 5 |
| | 2.2.1. Vineyards | 0.21% | 1.03 | 5 |
| | 2.2.2. Fruit trees and berry plantations | 0.80% | 3.84 | 5 |
| | 2.2.3. Olive groves | 2.62% | 12.58 | 13 |
| | 2.3.1. Pastures | 1.96% | 9.40 | 9 |
| | 2.4.2. Complex cultivation patterns | 4.60% | 22.04 | 22 |
| | 2.4.3. Land principally occupied by agriculture. with significant areas of natural vegetation | 9.11% | 43.68 | 44 |
| | | | *Sum Agriculture:* | *338* |
| Forest | 3.1.1. Broad-leaved forest | 13.76% | 65.94 | 66 |
| | 3.1.2. Coniferous forest | 7.94% | 38.05 | 38 |
| | 3.1.3. Mixed forest | 5.17% | 24.77 | 25 |
| | | | *Sum Forest:* | *129* |
| Water | 5.1.1. Water courses | 0.30% | 1.45 | 5 |
| | 5.1.2. Water bodies | 0.80% | 3.82 | 5 |
| | | | *Sum Water:* | *10* |
| | | *Total Sum:* 100.00% | *Total Sum:* | *539* |

2.3.3. Automated Class Label Retrieval

The sampling points that were randomly selected on the CLC2012 were used to retrieve automatically the corresponding values (labels) from the HRLs and GLC30 raster files. For this purpose a Matlab script was developed, which (i.) reads the table with the geographic location of the sampling points (X,Y), (ii.) reads the processed image files of HRLs and GLC30, (iii.) stores the value (label) of the pixel whose center is nearest to the given X, Y, and (iv.) creates a table with the sample id, the X, Y coordinates, and the corresponding labels from the HRLs and GLC30 layers. These files were further edited in order to contain also the LC labels in text (Table 4).

**Table 4.** An example with 10 arbitrary selected samples (id) from the CLC 2012 Level 3 1.1.1. Continuous Urban Fabric subclass, the corresponding LC values and classes for the HRLs and GLC30 layers.

| id | CORINE code/class | GLC30 pixel value | GLC30 Class | HRLs pixel value | HRLs class |
|---|---|---|---|---|---|
| 0 | 1.1.1 Continuous Urban Fabric | 80 | Artificial Surfaces | 92 | Imperviousness |
| 1 | 1.1.1 Continuous Urban Fabric | 80 | Artificial Surfaces | 89 | Imperviousness |
| 2 | 1.1.1 Continuous Urban Fabric | 80 | Artificial Surfaces | 99 | Imperviousness |
| 3 | 1.1.1 Continuous Urban Fabric | 80 | Artificial Surfaces | 97 | Imperviousness |
| 4 | 1.1.1 Continuous Urban Fabric | 80 | Artificial Surfaces | 74 | Imperviousness |
| 5 | 1.1.1 Continuous Urban Fabric | 80 | Artificial Surfaces | 96 | Imperviousness |
| 6 | 1.1.1 Continuous Urban Fabric | 80 | Artificial Surfaces | 90 | Imperviousness |
| 7 | 1.1.1 Continuous Urban Fabric | 80 | Artificial Surfaces | 0 | unclassified |
| 8 | 1.1.1 Continuous Urban Fabric | 80 | Artificial Surfaces | 65 | Imperviousness |
| 9 | 1.1.1 Continuous Urban Fabric | 80 | Artificial Surfaces | 99 | Imperviousness |
| 10 | 1.1.1 Continuous Urban Fabric | 80 | Artificial Surfaces | 100 | Imperviousness |

2.3.4. Reference Data Annotation based on Google Earth Images

The reference (Ground Truth) data were manually created after an intensive manual image interpretation procedure. For every sample two image interpretation experts assigned a LC label, i.e., *Artificial Surfaces, Forest, Water, Agriculture*, based on Google Earth very high resolution imaging data from the same year and a date, as close to the date of raw image acquisition as possible, for the generation of the C/GLCs. Especially for the *Water* class the former prerequisite was given special attention, and Google sightings both prior and after the raw image acquisition are taken into account. In particular, the procedure included the following steps: (i) the sampling points for each subclass (Level 3) of the CLC2012 were translated into kml files and then (ii) were inserted into Google Earth, so (ii) each expert assigned a LC label by interpreting the high resolution Google Earth images of the same date/period. Since reference imagery of higher resolution is used than the resolution of the maps assessed, a smaller MMU is possible for the reference labeling [22]. The aim here was to interpret directly on the sampling point over Google Earth imagery for the reference classification. However, for cases that one sample was on/near the borders of two or more different classes the experts examined the surrounding area and dominating classes, using a patch of about 60mx60m around the point in order to also account for the geolocation error observed between the different products (see Section 2.3.2).

At the same time the experts provided a confidence level indicating their scoring confidence for each and every annotation while generating the ground truth data base. Three confidence levels were assigned i.e., #1 for >75% confidence level, #2 for 25%-75% confidence level and #3 for confidence level below 25%. For all samples (approximately 250) that were annotated with a different label by the experts or were annotated with lower confidence levels (<75%), a second revision was performed by both experts in order to reach and agree to a common consensus.

2.3.5. Stratified sampling based on HRLs and GLC30.

In order to evaluate whether the stratified sampling procedure performed, based on a different LC product (*e.g.* HRLs, GLC30), may affect the Overall Accuracy (OA) scores, two additional sets of validation samples were generated based on the other two land cover products by employing again a stratified sampling methodology. Since HRLs and GLC30 products do not consist of more detailed subclasses as CLC2012, samples based on the relative coverage of the general classes *Artificial Surfaces*, *Forest*, *Water* and *Agriculture* were generated. The HRLs product does not provide an *Agriculture* layer; thus no validation samples could be generated for this class on the HRLs-based approach.

In order to compare the effect of the different sampling sets the number of samples for one of the classes (*i.e., Forest* with number of samples: 129) was kept the same. The number of the other classes' samples for both HRLs and GLC30 layers were calculated accordingly based on a stratified sampling procedure. Again a minimum of 5 samples was set for each category (Tables 5 and 6). Figure 2 presents the distribution of samples per category for the different stratified sampling procedures applied based on the three different datasets.

**Table 5.** The HRLs-based stratified sampling using the common in all experiments 129 number of samples for the class Forest.

| **HRLs-based Stratified Sampling** | | | |
|---|---|---|---|
| **Class** | **Area Coverage** | **# of samples per coverage** | **Selected # of samples** |
| Artificial Surfaces | 4.17% | 5.71 | 6 |
| Forest | 94.14% | 129.00 | 129 |
| Water | 1.69% | 2.32 | 5 |
| Sum: | 100.00% | | 140 |

**Table 6.** The GLC30-based stratified sampling using the common in all experiments 129 number of samples for the class Forest.

| **GLC30-based Stratified Sampling** | | | |
|---|---|---|---|
| **Class** | **Area Coverage** | **# of samples per coverage** | **Selected # of samples** |
| Artificial Surfaces | 3.36% | 11.33 | 11 |
| Forest | 38.23% | 129.00 | 129 |
| Water | 0.61% | 2.07 | 5 |
| Agriculture | 57.80% | 195.01 | 195 |
| Sum: | 100.00% | | 340 |

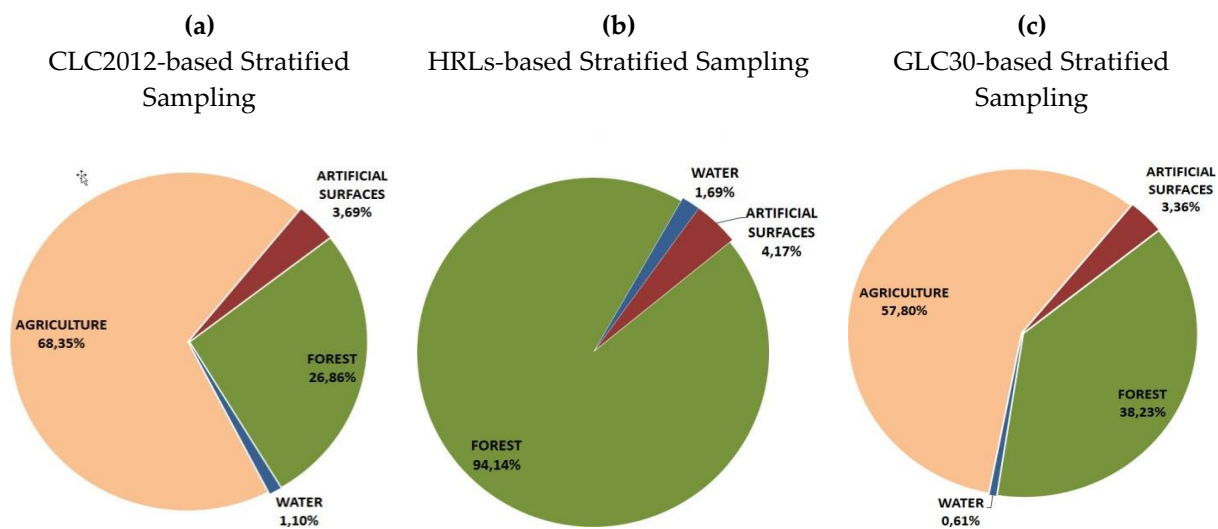

**Figure 2.** The proportions of samples per LC class that were derived based on a stratified sampling procedure applied on the three different datasets: (**a**) CLC2012, (**b**) HRLs and (**c**) GLC30.

2.3.6. Weighted accuracy assessment

The performance of each LC map was validated forming confusion matrices against the manual annotated GT data. Agreement between the maps and reference classifications was defined by the class correspondence shown in Table 1 and Table 2. In order to exploit the level of confidence assigned by the experts' interpretation different confusion matrices were computed per confidence level. The standard accuracy metrics of OA, User's and Producer's Accuracy (UA and PA) and kappa coefficient were calculated for each case. The incorporation of the confidence parameter to the validation process was achieved through equation 2, which combines the accuracy metric result to the respective weight of the level of confidence and number of observations in order to obtain a weighted overall accuracy:

$$wA = \frac{\sum_{i=1}^{3} w_i * N_i * A_i}{\sum_{i=1}^{3} w_i * N_i} \quad (2)$$

where $w_i$ is the given weight for the i confidence level, $N_i$ the number of observations and $A_i$ the achieved accuracy metric rate for the given confidence group.

**3. Experimental Results and Validation**

In this Section results from the validation of the three LC datasets against the manually annotated Ground Truth (GT) data are presented. In particular, in sub-section 3.1. results obtained based on validation samples from the L3 of CLC2012 are presented. This was the main focus of the present study since the availability of information regarding the area coverage of several L3 subclasses guarantees a more consistent stratified sampling procedure. Then in sub-section 3.2. the contribution of the confidence level that the experts assigned for each interpretation is analysed. In subsection 3.3. results from the validation and inter-comparison utilizing the HRLs' and GLC30's validation sampling datasets are demonstrated and commented.

*3.1. Validation against the GT with sampling sets from the CLC2012 L3.*

The three LC datasets were validated against the GT data based on a weighted accuracy assessment (see section 2.3.6). Table 7 presents the resulting weighted OA for the GLC30 product. Firstly, the median (*M*) of the confidence scoring level is defined based on the percentage range and then the corresponding weight $w_i$ for each confidence class is calculated based on *M*. The OA between the GT data and the CLC2012 L3 sub-classes for confidence levels 1, 2 and 3 was respectively 90%, 78%, and 77%. The weighted OA based on Eq. 2 was calculated at 86%.

Table 7. The OA of the GLC30 per confidence level and the corresponding weighted OA

| **GLC30 | CLC2012-based sampling | Confidence Level: #1** | | | | |
|---|---|---|---|---|
| *Sum* | *M* | $w_i$ | #N of samples | Overall Accuracy |
| 1 (>75%) | 87.50 | 0.583 | 291 | 90% |
| 2 (25%-75%) | 50.00 | 0.333 | 218 | 78% |
| 3 (<25%) | 12.50 | 0.083 | 30 | 77% |
| Sum: | | 1 | 539 | **Weighted OA: 86%** |

Table 8. The confusion matrix for CLC2012 product for samples with high confidence level (CL #1).

| **CLC2012 | CLC2012-based sampling | Confidence Level: #1** | | | | | | |
|---|---|---|---|---|---|---|---|
| Ground Truth | **Artificial Surfaces** | **Agriculture** | **Forest** | **Water** | **Others/Unclassified** | *Sum* | **PA** |
| **Artificial Sur.** | **33** | 4 | 0 | 0 | 0 | 37 | 89% |
| **Agriculture** | 1 | **164** | 0 | 0 | 0 | 165 | 99% |
| **Forest** | 0 | 0 | **70** | 0 | 0 | 70 | 100% |
| **Water** | 0 | 0 | 0 | **7** | 0 | 7 | 100% |
| **Others/Unclassified** | 2 | 3 | 3 | 2 | **0** | 10 | 0% |
| *Sum* | 36 | 171 | 73 | 9 | 0 | 289 | |
| **UA** | 92% | 96% | 96% | 78% | - | | 95% |

Overall accuracy = **94.81%**

kappa = **0.91**

Table 8 presents the confusion matrix for the CLC2012 product and for the samples that were annotated with high confidence level #1. The OA reached a 95% with a 0.91 kappa coefficient. Four out of 37 *Artificial Surfaces* testing samples were falsely labelled as *Agriculture*, while one out of 165 of *Agriculture* was labelled as *Artificial Surfaces.* All GT validation samples for classes *Forest* and *Water* were successfully classified resulting to 100% PA. In general, all classes presented high levels of PA and UA, apart from class *Others/Unclassified,* which, as expected, presented zero classification cases to the corresponding GT class, since the validation is conducted for the CLC2012 using the CLC2012-based sampling dataset. The OA for CL: #2 reached 77% and the OA for CL: #3 was 72%.

By integrating all confidence levels together, the resulting weighted OA reached the 90% for the HRLs, the 89% for the CLC2012 and the 86% for the GLC30. Moreover, in Figure 3 one can observe that in all cases the *Agriculture* and *Forest* classes had PA and UA of above 85%, indicating high accuracy and reliability respectively. For CLC2012 the PA for *Artificial Surfaces* was 85%, for *Water* 100%; however, lower rates were recorded for the UA, *i.e.* 67% and 79%, respectively. GLC30 presented relatively high accuracy for the class *Artificial Surfaces* both for PA (74%) and UA (75%) metrics. However, GLC30 presented mis-classification cases for the class *Water*; the majority of GT samples (six out of eight) were classified as *Agriculture*, resulting into a weighted PA of 26%. Several GT samples of *Artificial Surfaces* were also mis-classified as *Others/Unclassified* in HRLs resulting in a relatively low PA rate of 55%. The class *Water* resulted into a 100% UA rate and a PA rate of 74% in the HRLs.

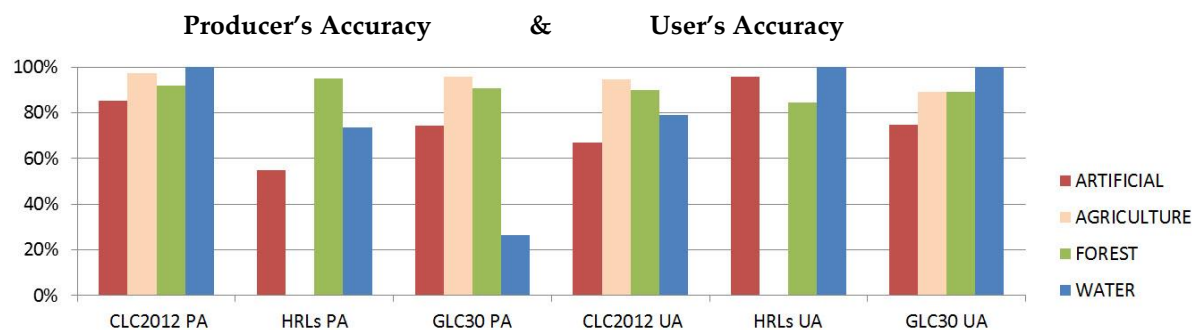

**Figure 3.** Weighted PA and UA for all studied LC datasets.

In Table 9 the highest weighed scores recorded for PA and UA rates are presented per LC class. As it can be observed, in five out of eight cases the CLC2012 product validation achieved the highest values.

**Table 9.** The highest weighted PA and UA rates per LC class.

| LC Class | PA | UA |
|---|---|---|
| Artificial Surfaces | ClC2012 (85%) | HRLs (96%) |
| Agriculture | ClC2012 (97%) | ClC2012 (95%) |
| Forest | HRLs (95%) | ClC2012 (90%) |
| Water | ClC2012 (100%) | HRLs & GLC30 (100%) |

*3.2. The contribution of the confidence indicator*

In this subsection the OA (and weighted OA) rates are compared both without and by taking into account the confidence level (CL) that the expert assigned at every GT point during the image interpretation. In Table 10, the OA rates per CL for all LC maps are presented along with the number of samples per CL. The last two rows present the OA and the weighted OA rates. The latter are increased by 1-3% compared to the OA ones. In particular, as the CL decreases the OA rate decreases, too. This is quite expected, since the annotations with lower confidence levels, indicating difficulty in the labelling decision, usually involve particular regions, terrain types and complex land cover/use cases, lying most probably on the borders of two LC classes, and therefore they are more associated with classification errors.

Similar conclusions are derived when comparing the resulting PA and UA per class rates for all products validation. In Table 11 one can also observe the differences between the standard and weighted PA and UA. For most cases the weighted PA and UA are increased around 1%-4%. Classes *Water* and *Agriculture* present the smallest differences. A greater difference occurs for the UA of CLC2012 product validation of *Artificial Surfaces*, which presents a UA rate of 58% and a weighted UA rate of 67%. This 9% difference occurs due to the fact that a rather large number of *Agriculture* and *Forest* samples in the GT data sets, characterized with a CL of 2, were annotated as *Artificial Surfaces* in the CLC2012 map. The contribution of these errors decreased when the weighted UA metric was calculated, since these CL2 observations are given a decreased weight in the calculation.

Table 10. Overall Accuracy and weighted Overall Accuracy rates for the CLC2012-based stratified sampling.

| LC Map | ClC2012 | | | HRLs | | | GLC30 | | |
|---|---|---|---|---|---|---|---|---|---|
| *Confidence Level* | *CL1* | *CL2* | *CL3* | *CL1* | *CL2* | *CL3* | *CL1* | *CL2* | *CL3* |
| *# of samples* | *289* | *225* | *25* | *289* | *225* | *25* | *291* | *218* | *30* |
| *OA per CL* | *95%* | *77%* | *72%* | *91%* | *87%* | *76%* | *90%* | *78%* | *77%* |
| **OA** | **86%** | | | **89%** | | | **84%** | | |
| **Weighted OA** | **89%** | | | **90%** | | | **86%** | | |

Table 11. The difference (%) between the standard PA, UA) and weighted PA and UA (wPA, WUA) rates.

| LC Maps | ClC2012 | | HRLs | | GLC30 | |
|---|---|---|---|---|---|---|
| | wPA - PA | wUA - UA | wPA - PA | wUA – UA | wPA – PA | wUA - UA |
| **Artificial Sur.** | 4% | 9% | 1% | 2% | 0% | -1% |
| **Agriculture** | 1% | 1% | - | - | 0% | 1% |
| **Forest** | 4% | 4% | 3% | 2% | 4% | 6% |
| **Water** | 0% | -1% | -1% | 0% | 1% | 0% |

In overall, the weighted OA, PA and UA result into a more clear estimation of the actual accuracies, since they integrate a measure of confidence, especially for the GT samples that are vague or borderline.

*3.3. Validation based on a sampling design from the HRLs and GLC30*

In order to evaluate if and in which extent the selection of the product, on which the stratified sampling is based, influences the resulting OA two extra sampling datasets are incorporated into the analysis (see Section 2.3.5.). In Table 12 the weighted Overall Accuracy results for the validation of all land cover products using all sampling datasets, are presented. It is observed that in every case the weighted OA metric reached high rates of above 84%. The most stable OA rates were derived from the CLC2012 and GLC30 datasets, when different GT data sets were employed. The HRLs OA rate falls below 90% (at 84%), when the stratified sampling is based on its own layer. However, at the same time the HRLs product scored the highest weighted OA rates i.e., 90%, 93% for stratified sampling based on CLC2012 and GLC30, while the CLC2012 followed with 89% and 87% respectively. The GLC30 product performed slightly lower rates of 86% and 84%. The HRLs dataset included validation samples solely from the classes *Artificial Surfaces, Forest* and *Water,* omitting the non-present *Agriculture* class. As a consequence, the outcome pattern of the results differs a bit from the other two datasets. Still the accuracy rates are really high; for HRLs-based sampling CLC2012 product holds the best OA rate with 89%, the GLC30 follows with 85% and the HRLs with 84%.

Table 12. Weighted OA from three different stratified sampling procedures.

| | **Weighted OA** | | |
|---|---|---|---|
| | **ClC2012-based** | **HRLs-based** | **GLC30-based** |
| **Sampling dataset** | Classes: *Artificial Surfaces, Agriculture, Forest, Water, Others/Unclassified* | Classes: *Artificial Surfaces, Forest, Water, Others/Unclassified* | Classes: *Artificial Surfaces, Agriculture, Forest, Water, Others/Unclassified* |
| | 539 samples | 140 samples | 340 samples |
| **CLC2012** | 89% | 89% | 87% |
| **HRLs** | 90% | 84% | 93% |
| **GLC30** | 86% | 85% | 84% |

The weighted PA and UA are also presented in Figure 4. By comparing also with the CLC2012-based results (Figure 3) one can observe that *Artificial Surfaces* present the same error pattern regarding PA across the different sampling datasets. More analytically for this class, CLC2012 delivered the highest PA rates (*i.e.,* 85%, 99%, 100%) and HRLs the lowest ones (*i.e.,* 55%, 54%, 62%), based on CLC2012, GLC30 and HRLs sampling respectively. In all cases, the *Agriculture* and *Forest* classes retained a PA above 90%. The *Water* class delivered a significantly low PA (<30%) for the GLC30 product in all sampling datasets, while for the CLC2012 product the same rates were above 80% (and up to 100% for CLC2012 and GLC30 based sampling). For HRLs product validation of the *Water* class rates exceeded 74% and reached 100% when validating with HRLs based sampling dataset.

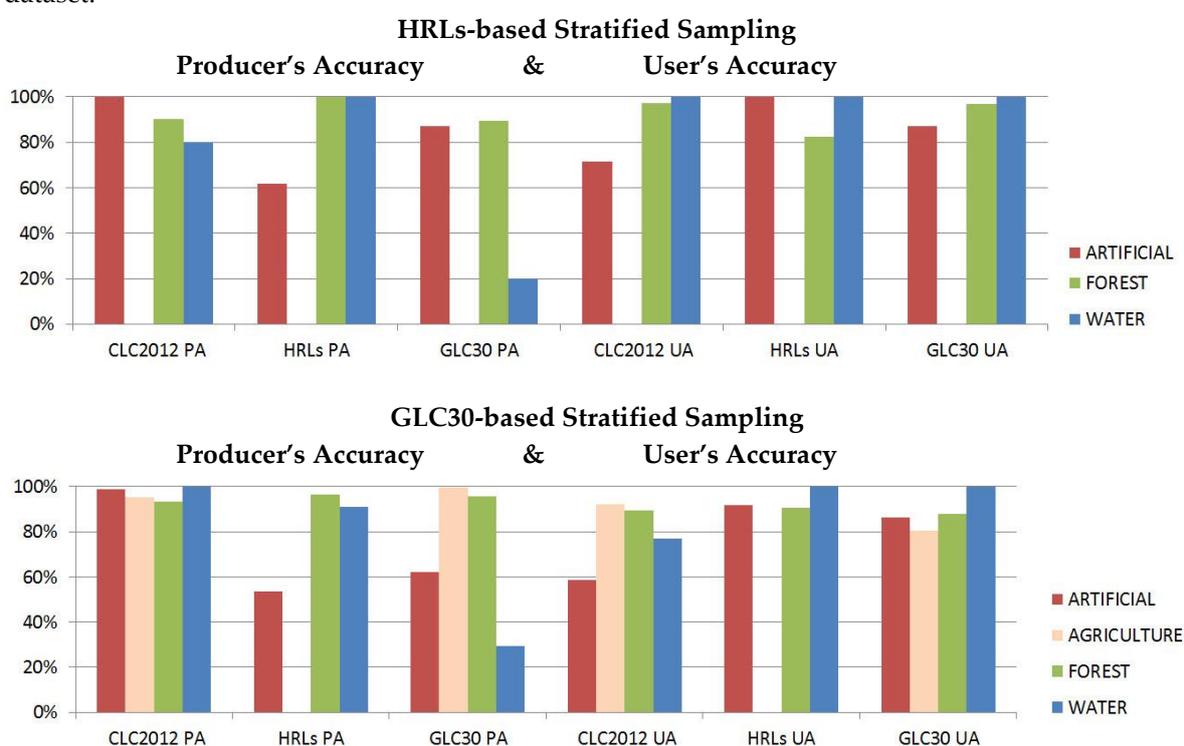

**Figure 4.** Weighted PA and UA from a stratified sampling based on the HRLs and GLC30 datasets.

Regarding the UA rates, *Artificial Surfaces* present again the same pattern across the different sampling datasets however in a reverse order than PA. In particular, for this class the HRLs scored the highest UA rates (96%, 92%, 100%) and ClC2012 the lowest ones (67%, 59%, 71%), based on CLC2012, GLC30 and HRLs sampling respectively. In all cases, the *Agriculture* and *Forest* classes delivered high UA rates (>80%). The *Water* class UA was in most cases at 100% apart from the ClC2012 product based on sampling sets from ClC2012 (79%) and GLC30 (77%). Moreover, as it can be observed in Table 13 the CLC2012 product resulted in most cases with the highest weighted PA and UA rates (5 out 8 experiments) with different sampling sets. The HRLs are, also, resulting with high accuracy rates; however, in most cases with sampling sets derived from their own stratified sampling procedure.

Table 13. The best and worst cases for the weighted PA and UA rates from all experiments (based on various GT data sets).

| Class | Producer's Accuracy | | User's Accuracy | |
| --- | --- | --- | --- | --- |
| | Best | Worst | Best | Worst |
| *Artificial Surfaces* | **ClC2012** (HRLs-based) 100% | HRLs (GLC30-based) 54% | **HRLs** (HRLs-based) 100% | ClC2012 (GLC30-based) 59% |
| *Agriculture* | **GLC30** (GLC30-based) 100% | ClC2012 (GLC30-based) 95% | **ClC2012** (ClC2012-based) 95% | GLC30 (GLC30-based) 80% |
| *Forest* | **HRLs** (HRLs-based) 100% | ClC2012 & GLC30 (HRLs-based) 90% | **ClC2012 & GLC30** (HRLs-based) 97% | HRLs (HRLs-based) 82% |
| *Water* | **ClC2012** (ClC2012-based & GLC30-based) **HRLs** (HRLs-based) 100% | GLC30 (HRLs-based) 20% | **All** (100%) apart from ClC2012 with ClC2012-based & GLC30-based SS | ClC2012 (GLC30-based) 77% |

## 4. Discussion

Although various validation frameworks for the qualitative and quantitative assessment of C/GLCs have been presented in several studies [23-28] they neither used a common reference layer for the inter-comparison nor did they incorporated a confidence level during the reference data production. Based on (i) the confidence levels that were recorded during the reference/ground truth data production, and on (ii) the performed stratified sampling on the L3 CLC classes, the resulting weighted overall accuracy rates per land cover product indicate and highlight more objectively further qualitative and quantitative aspects of the considered C/GLCs. In particular, the reported quantitative results are similar with the ones presented in [24, 25] and disagree with the relative lower OA rates (<50%) presented in [28]. The quantitative results from the assessment in [28] disagree with most of the literature. It should be noted that in the present study the *Bareland* class was not evaluated, while in [28] *Bareland* was the one with the most mis-classification errors. Qualitatively in all considered C/GLCs, *Artificial Surfaces* is the class with the more commission errors, resulting into lower UA rates, while *Water* is the one with usually the highest reliability.

**Sampling design strategy**

In contrast to similar efforts in the literature (see Section 1. Introduction), here it is examined whether the selection of sample datasets from different LC products is affecting the reported OA rates. In particular, the collection of validation samples was initially designed to follow the stratified approach based on the CLC2012 L3 subclasses and the calculated extent of each subclass in the study area, setting a maximum of 120 samples and a minimum of 5 samples. The initial goal was to annotate approximately 500 samples. However, since the sampling design procedure significantly affects the determination of the thematic map accuracy estimation [29], two extra sample datasets were generated based on the nomenclature and classification scheme of the other two land cover maps. These extra validation datasets (sampling sets based on HRLs and GLC30) were designed in order to examine any sort of biases in class representation. The reported OA accuracy from all experiments with the three different sampling sets (see Table 12) did not indicate significant differences i.e., less than 3% apart from the validation of HRLs under a sampling set derived from its own layer. It should be also noted that in all cases the evaluation based on HRLs did not include the challenging *Agriculture* class.

**Quantify confidence levels during GT annotation and Impact on Results**

In this work the reference classification was based on reference imagery of higher resolution than those of the map assessed and was also intended to apply to a variety of maps, so the MMU differed from the maps' spatial assessment units [22]. In particular interpretation was applied directly on the sampling points over Google Earth imagery for the reference classification, though for vague cases we applied a patch of about 60mx60m around the point in order to also account for the geolocation error observed between the different products. Furthermore, in this study during image interpretation and annotation procedure, experts registered their confidence level for each and every sampling point. This has been already proposed and highlighted as a good practice for accuracy assessment in the literature towards a way to characterize interpreter's perception of uncertainty in the reference classification [3, 22, 29].

In accordance with [30] in this work a nominal confidence level in the labels of chosen reference samples is given. Low, moderate and high confidence rates were used in the analysis in order to subset the results by confidence [17, 30], i.e., the following ratings were used: #1 for >75% confidence, #2 for 25%-75% confidence, and #3 for <25% confidence. In accordance with the literature, the visual interpretation, here, has been proven to be highly biased depending on the interpreter [3, 29]. Therefore, for every sample interpreted with lower confidence level or for cases that the two experts disagreed on labelling, a second round of interpretation took place in order to reach consensus.

The OA rates without integrating the confidence levels were in general lower than the weighed OA rates, when the assigned confidence levels were integrated in the evaluation. The higher weighed OA rates derived from the fact that in most cases the mis-classification errors were for sampling cases that the experts, also, hardly could reach a decision due to terrain, land cover/use complexity, borderline cases, etc. It should be noted, also, that most of the automatically and randomly selected GT samples were annotated with high or medium confidence levels and only a 5% of the samples were annotated with the lower confidence level #3.

Since, similar studies have indicated that OA results from a confusion matrix should be interpreted with caution, as the matrix records the degree of agreement between the reference data and the map data, which are in cases less than perfect [17, 31], it is suggested here that the integration of confidence levels during annotation can well address these concerns.

**5. Conclusions**

The validation procedure proposed in this study provides a qualitative and statistical outcome comparing the accuracy of three C/GLC products (namely CLC 2012, HRLs and Globeland30) in detecting *Artificial*, *Forest*, *Agriculture* and *Water* classes for a study area in central Greece, Thessaly. In particular, based on all performed experiments, the HRLs and the CLC2012 resulted into the highest overall accuracy rates with an average OA of 89% and 88%, respectively. However, the evaluation of CLC2012 included also the *Agriculture* class, which is an important and challenging LC category. The GLC30 resulted into an average OA of 85%. Latter slight lower performance can be attributed to misclassifications for the classes *Water* and *Artificial Surfaces*.

The experts' confidence level during GT annotation was recorded and integrated in the evaluation process through a weighted accuracy assessment procedure, while different sampling sets based on all 3 LC products were juxtaposed for the evaluation. In all cases overall weighted accuracy rates exceeded 80%. The main conclusion outlines that the proposed methodology and analysis, manages to highlight and indicate more details of the quantitative and qualitative aspects of the considered LC maps and enhances their credibility to the users.

**Acknowledgments:** We gratefully acknowledge support from the Research Projects for Excellence IKY (State Scholarships Foundation) / SIEMENS. Authors would like to acknowledge that the presented work is generated in support of the activities for the "Global Land Cover Products Validation and Inter-comparison in South Central and Eastern Europe" project within the South Central and Eastern European Regional Information Network (SCERIN)

**Author Contributions:** I.M. and K.K. conceived and designed the experiments; X.K. and G.G. performed the experiments; X.K. and G.G. analyzed the data; I.M. and K.K. contributed materials/analysis tools; I.M., X.K., K.K. and G.G. wrote and edited the manuscript.

**Conflicts of Interest:** The authors declare no conflict of interest.